\documentclass[prd,showpacs,amsmath,amssymb,superscriptaddress,preprintnumbers,nofootinbib,showkeys]{revtex4-2}
\usepackage{graphicx}

\begin{document}

\title{Nonlinear axion electrodynamics: \\ Axionically induced electric flares in the early  magnetized universe}

\author{Alexander B. Balakin}
\email{Alexander.Balakin@kpfu.ru} \affiliation{Department of
General Relativity and Gravitation, Institute of Physics, Kazan
Federal University, Kremlevskaya str. 16a, Kazan 420008, Russia}

\author{Vladimir V. Bochkarev}
\email{Vladimir.Bochkarev@kpfu.ru} \affiliation{Department of
Radiophysics, Institute of Physics, Kazan
Federal University, Kremlevskaya str. 16a, Kazan 420008, Russia}

\author{Albina F. Nizamieva}
\email{alb9061@yandex.ru} \affiliation{Department of
General Relativity and Gravitation, Institute of Physics, Kazan
Federal University, Kremlevskaya str. 16a, Kazan 420008, Russia}

\date{\today}

\begin{abstract}
We consider the nonlinearly extended Einstein-Maxwell-axion theory, which is based on the account for two symmetries: first, the discrete symmetry associated with the properties
of the axion field, second,  the Jackson's symmetry, prescribing to the electrodynamics to be invariant with respect to the rotation in the plane coordinated by the electric and
magnetic fields.  We derive the master equations of the nonlinearly extended theory, and apply them to the Bianchi-I model with magnetic field. The main result, describing the behavior
of the nonlinearly coupled axion, electromagnetic and gravitational fields is the anomalous growth of the axionically induced electric field in the early magnetized Universe.
The character of behavior of this anomalous electric field can be indicated by the term flare. We expect, that these electric flares can produce the electron-positron pair creation,
significant acceleration of the born charged particles, and emission of the electromagnetic waves by these accelerated particles.
\end{abstract}
\maketitle

\section{Introduction}

The concept of Symmetry has played and continues to play outstanding role in Physics of Fundamental Interactions. Extending the theoretical models of the Nonlinear cosmic axion electrodynamics, we happen to be  inspired by beauty of the symmetries of two types.
The first type of symmetry is connected with the idea that the Lagrangian of the axion electrodynamics is invariant with respect to the discrete symmetry, associated with the shift $\tilde{\phi} = \phi {+} 2\pi k$ ($k$ is an integer) of the pseudoscalar field $\phi$ introduced by Peccei and Quinn \cite{PQ} and indicated later as axion field (see, e.g. \cite{1,2,3,4,5} for details of this story). On this way we need to have a Lagrangian, which contains the invariants based on the Maxwell tensor $F_{ik}$, its dual $F^*_{ik}$, and the axion field $\phi$ introduced nonlinearly. Clearly, the well-known linear term $\phi F^{*}_{mn}F^{mn}$ supports the discrete symmetry, since the rest term $2\pi k \sqrt{-g} F^{*}_{mn}F^{mn}$ in the action functional is the perfect divergence and thus can be eliminated. When one deals with the nonlinear modification of the theory, the nonlinear function of the true invariant $f\left(\phi F^{*}_{mn}F^{mn}\right)$ does not satisfy the mentioned discrete symmetry, and in this sense we have to search for new nonlinear terms. For this purpose we attract for consideration the idea of Jackson \cite{Jackson} that the vacuum Faraday-Maxwell electrodynamics is symmetric with respect to linear transformation of the electric and magnetic fields ${\vec{E}}_* \to \vec{E} \cos{\alpha} {+} \vec{B} \sin{\alpha}$, ${\vec{B}}_* \to \vec{B} \cos{\alpha} {-} \vec{E} \sin{\alpha}$, with the constant angle $\alpha$. On the way of nonlinear modification of the axion electrodynamics we assume to use the Lagrangian, which satisfies both mentioned symmetries. The example of such Lagrangian has been constructed in the  work \cite{BaGa}. The proposed nonlinear Lagrangian has the form ${\cal L}({\cal I})$, where ${\cal L}({\cal I})$ is some nonlinear function of its argument, and  the invariant ${\cal I}$ is
\begin{equation}
{\cal I}= \frac14 \left(\cos{\phi} \ F_{mn}F^{mn}+ \sin{\phi} \ F^{*}_{mn}F^{mn}\right) \,.
\label{01}
\end{equation}
We assume that for small ${\cal I}$ the Lagrangian can be transformed into ${\cal L}({\cal I}) \to {\cal I}$. In the linear theory, when $\phi \to 0$, the invariant ${\cal I}$ can be reduced to the standard invariant
\begin{equation}
{\cal I} \to  \frac14 \left(F_{mn}F^{mn}+ \phi F^{*}_{mn}F^{mn}\right)
\label{02}
\end{equation}
linear in $\phi$ and covering the basic term of the linear axion electrodynamics \cite{3,4}. Clearly, the discrete symmetry $\tilde{\phi} = \phi + 2\pi k$ is supported identically by the choice of sin/cosine multipliers in front of the standard terms $F_{mn}F^{mn}$ and $F^{*}_{mn}F^{mn}$. As for the Jackson's symmetry,
if we use the transformation
\begin{equation}
{\cal F}^{ik} =  F^{ik}\cos{\frac{\phi}{2}} +  F^{*ik} \sin{\frac{\phi}{2}} \,,
\label{5}
\end{equation}
we obtain that the term $\frac14 {\cal F}^{ik}{\cal F}_{ik}$
coincides with (\ref{01}). This fact confirms the validity of the symmetry transformations
\begin{equation}
{\cal E}^i =  E^{i} \cos{\frac{\phi}{2}} +  B^{i} \sin{\frac{\phi}{2}} \,, \quad {\cal B}^i =  - E^{i} \sin{\frac{\phi}{2}}  + B^{i} \cos{\frac{\phi}{2}} \,,
\label{4}
\end{equation}
which include the electric field four-vector $ E^i = F^{ik}U_k$ and the magnetic induction four-vector $B^i=F^{*ik}U_k$. The quantity $U^k$ is the velocity four-vector of the medium or of an observer.

The nonlinear models with such symmetries attract special attention, since they contain a latent instability which can be displayed by the anomalous growth of the axionically induced electric field. When the model is quasi-linear, i.e., when the axion field enters the Lagrangian with sin/cosine nonlinearities, but the electromagnetic field is described in the linear context, the problem of electric flares has been discussed in the paper \cite{BaGa}. It was shown in \cite{BaGa} that the interaction with the dynamic aether stabilizes the evolution, and the axion field does not reach the catastrophic value $\phi = \frac{\pi}{2}$, which corresponds to the infinite value of the electric field proportional to the $\tan{\phi}$. Now we exclude the dynamic aether from the model, but consider the electromagnetic field in the nonlinear context. Our goal is to show that in the manifold of the guiding parameters of the model there exists an area, in which the presented model is stable. In particular, this means that the axionically induced electric field is finite, but its amplitude can reach very large values thus producing the pair creation, particle acceleration and electromagnetic waves emission in the early magnetized Universe.

The paper is organized as follows. In Section II we describe the mathematical formalism of the proposed nonlinear extension of the Einstein-Maxwell-axion theory. In Section III  we reduce the basic equations taking into account the symmetry of the cosmological anisotropic Bianchi-I model with magnetic field, and derive the pair of key equations for the axion field and for the derivative of the scale factor. In Section IV we analyze the solutions to the key equations using qualitative and numerical methods, and discuss the problem of the axionically induced electric field generation.
Section V contains discussion and conclusions.

\section{Nonlinear Einstein-Maxwell-Axion Model}

\subsection{The Action Functional and Master Equations of the Model}

The model is based on the action functional presented in the form
\begin{equation}
- S = \int d^4x \sqrt{-g} \left\{\frac{R{+}2\Lambda}{2\kappa} {+} \frac12 \Psi^2_0  \left[V(\phi) {-} \nabla_k \phi \nabla^k \phi \right]{+} {\cal L}({\cal I}) \right\}
\,.
\label{act}
\end{equation}
Standardly, $R$ is the Ricci scalar, $\Lambda$ is the cosmological constant, $\kappa = 8 \pi G$ is the Einstein constant (we consider $c=1$). $\phi$ describes the pseudoscalar (axion) field; $V(\phi)$ is the potential of the axion field, $\nabla_k$ is the covariant derivative, and the constant $\Psi_0$ is reciprocal to the coupling constant of the axion-photon interaction $g_{A \gamma \gamma}$, i.e.,  $g_{A \gamma \gamma}= \frac{1}{\Psi_0}$. The new element of the theory is encoded in the last term ${\cal L}({\cal I})$, the Lagrangian of the electromagnetic field interacting with the axion field; this Lagrangian is the nonlinear function of the argument ${\cal I}$, which has at least one continuous derivative and tends to ${\cal I}$ at ${\cal I} \to 0$, i.e., ${\cal L}({\cal I} \to 0)  \to {\cal I}$. The argument ${\cal I}$ is presented by (\ref{01}).

\subsubsection{Master Equations for the Electromagnetic Field}

Variation of the functional (\ref{act}) with respect to the potential of the electromagnetic field $A_i$ yields
\begin{equation}
\nabla_k \left\{ {\cal L}^{\prime}({\cal I})\left[\cos{\phi} \ F^{ik} + \sin{\phi} \ F^{*ik} \right] \right\}=0 \,,
\label{eld}
\end{equation}
where the prime denotes the derivative with respect to the argument ${\cal I}$. In addition to the nonlinear electrodynamic equation (\ref{eld}) we use the standard equations
\begin{equation}
\nabla_k F^{*ik}=0 \,,
\label{11}
\end{equation}
which are the consequence of the definition of the Maxwell tensor $F_{mn}= \nabla_m A_n {-} \nabla_n A_m$.

As examples, below we consider two model variants of the nonlinear function  ${\cal L}({\cal I})$. The first one is of the power law type
\begin{equation}
{\cal L}({\cal I}) =  {\cal I} + \gamma {\cal I}^{\nu} \,, \quad \nu > 1 \,.
\label{L1}
\end{equation}
The second variant contains the Kohlrausch type function (stretched exponential \cite{Kohl})
\begin{equation}
{\cal L}({\cal I}) =  {\cal I}^{1-\gamma} \left[\exp\left({\cal I}^{\gamma}\right) -1 \right]\,, \quad \gamma >0\,.
\label{L3}
\end{equation}

\subsubsection{Master Equation for the Axion Field}

Variation of the functional (\ref{act}) with respect to the axion field $\phi$ gives the equation
\begin{equation}
\nabla^k \nabla_k \phi + \frac12 \frac{dV}{d\phi} =  \frac{1}{4\Psi^2_0}  {\cal L}^{\prime}({\cal I}) \left[\sin{\phi}  F_{ik}F^{ik} {-} \cos{\phi}   F_{ik}F^{*ik} \right] \,.
\label{13}
\end{equation}
We consider the potential $V(\phi)$ to be of the periodic form
\begin{equation}
V(\phi) = 2 m^2_A \left(1-\cos{\phi} \right) \,.
\label{14}
\end{equation}
For small values $\phi$ the potential (\ref{14}) converts into the standard one $V(\phi) \to m^2_A \phi^2$. Taking into account (\ref{14}), one can rewrite the equation (\ref{13}) as follows
\begin{equation}
\nabla^k \nabla_k \phi = - m^2_A  \sin{\phi} + \frac{1}{4\Psi^2_0}  {\cal L}^{\prime}({\cal I}) \left[\sin{\phi}  F_{ik}F^{ik} {-} \cos{\phi}   F_{ik}F^{*ik} \right] \,.
\label{139}
\end{equation}
Clearly, this equation satisfies explicitly the requirement of the discrete symmetry $\tilde{\phi} = \phi {+} 2\pi k$. Mention should be made that for the linear version of electrodynamics and for small $\phi$ the axion field equation
\begin{equation}
\nabla^k \nabla_k \phi +   \phi \left(m^2_A - \frac{1}{4\Psi^2_0} F_{ik}F^{ik}\right) = - \frac{1}{4\Psi^2_0}  F_{ik}F^{*ik}
\label{137}
\end{equation}
contains the supplementary term, which introduces the effective mass $M_{(\rm eff)}$ given by
\begin{equation}
M_{(\rm eff)} = \sqrt{m^2_A - \frac{1}{4\Psi^2_0} F_{ik}F^{ik}} \,,
\label{139}
\end{equation}
This situation is typical for the models of the dilaton-photon interactions, and this question has been discussed in \cite{BaGa}.

\subsubsection{Master Equations for the Gravitational Field}

Variation of the functional (\ref{act}) with respect to the metric gives the gravity field equations
\begin{equation}
R_{ik} - \frac12 R g_{ik} = \Lambda g_{ik} + \kappa T^{(\rm EMA)}_{ik} + \kappa T^{(\rm A)}_{ik} \,.
\label{15}
\end{equation}
Here the term
\begin{equation}
T^{(\rm A)}_{ik}= \Psi^2_0 \left[\nabla_i \phi \nabla_k \phi  + \frac12 g_{ik}\left(V - \nabla_p \phi \nabla^p \phi \right)\right]
\label{16}
\end{equation}
is the stress-energy tensor of the pseudoscalar (axion) field, and the tensor
\begin{equation}
T^{(\rm EMA)}_{ik} = {\cal L}^{\prime}({\cal I}) \cos{\phi} \left[\frac14 g_{ik} F_{mn}F^{mn} - F_{im}F_k^{\ m}\right] + g_{ik} \left[{\cal L}({\cal I}) - {\cal I} \cdot {\cal L}^{\prime}({\cal I}) \right]
\label{17}
\end{equation}
contains all the contributions of the electromagnetic field. It coincides with the standard stress-energy tensor of the electromagnetic field, when $\phi=0$ and ${\cal L}({\cal I})= {\cal I}$.
The trace of the tensor (\ref{17})
\begin{equation}
T^{(\rm EMA)}_{ik} g^{ik} = 4 \left[{\cal L}({\cal I}) - {\cal I} \cdot {\cal L}^{\prime}({\cal I}) \right]
\label{170}
\end{equation}
is generally non-vanishing; it is equal to zero only in the linear version of the theory.
The total stress-energy tensor satisfies the condition
\begin{equation}
\nabla^k \left[T^{(\rm A)}_{ik} + T^{(\rm EMA)}_{ik} \right]= 0
\label{179}
\end{equation}
as the differential consequence of (\ref{eld}), (\ref{11}), (\ref{13}).

\section{Magnetized Universe: Bianchi-I Model}

We apply the established theory to the anisotropic homogeneous cosmological model with magnetic field; the appropriate spacetime platform for this model is the  Bianchi-I model  (see, e.g. \cite{M1,M2,M3,M4,M5,M6} and references therein). The metric
\begin{equation}
ds^2 = dt^2 - a^2(t)dx^2 - b^2(t)dy^2 -c^2(t)dz^2
\label{18}
\end{equation}
depends on the cosmological time only. We assume that all physical quantities inherit this symmetry and also are the function of $t$ only.

\subsection{Exact Solution to the Electromagnetic Field Equations}

With the fixed symmetry ansatz, the equations  (\ref{11}) standardly give
the solution for the magnetic field
\begin{equation}
F_{12} = const \equiv B_0 \,,
\label{19}
\end{equation}
and hint that the electric field parallel to the magnetic field $E(t) \equiv c(t) F^{30}$ appears as the result of the axion - photon coupling.
Indeed, four equations (\ref{eld}) reduce to the following one:
\begin{equation}
\frac{d}{dt} \left\{ {\cal L}^{\prime}({\cal I}) \left[a(t)b(t) \cos{\phi(t)} E(t) - B_0 \sin{\phi(t)} \right]\right\}=0 \,,
\label{20}
\end{equation}
and its solution is
\begin{equation}
E(t) = \frac{B_0}{a(t)b(t)} \tan{\phi} + \frac{\rm const}{ab \cos{\phi} \ {\cal L}^{\prime}({\cal I})} \,.
\label{21}
\end{equation}
For the sake of simplicity we consider the model, in which the electric field is equal to zero at the moment $t_0$; also we assume that the initial value of the axion field is equal to $\pi m$ ($m=0,1,...$), thus providing that the integration constant is equal to zero
\begin{equation}
E(t_0)=0  \,, \quad \phi(t_0)= \pi m  \ \to  \ {\rm const} = 0 \,.
\label{22}
\end{equation}
With these initial conditions we obtain the elegant formulas, first,  for the electric field
\begin{equation}
E(t) =  \frac{B_0}{a(t)b(t)} \tan{\phi(t)} \,,
\label{23}
\end{equation}
second, for the invariants of the electromagnetic field
\begin{equation}
F^*_{pq}F^{pq} = \frac{4B_0^2}{a^2 b^2} \tan{\phi} \,, \quad F_{pq}F^{pq} =
\frac{2B_0^2}{a^2 b^2}\left[1- \tan^2{\phi}\right] \,,
\label{42}
\end{equation}
third, for the invariant ${\cal I}$
\begin{equation}
{\cal I} = \frac{B_0^2}{2a^2 b^2 \cos{\phi}} \,.
\label{43}
\end{equation}

\subsection{Reduced Equation for the Axion Field}

Equation (\ref{13}) with (\ref{14}) can be now transformed into
\begin{equation}
\ddot{\phi} + \frac{(abc)^{\cdot}}{abc} \dot{\phi} + m^2_{A} \sin{\phi}
= - \frac{B^2_0 {\cal L}^{\prime}({\cal I}) \sin{\phi}}{2\Psi^2_0 a^2 b^2 \cos^2{\phi}} \,,
\label{24}
\end{equation}
where the dot denotes the derivative with respect to the cosmological time. There is also the following convenient form of this equation
\begin{equation}
\ddot{\phi} + \frac{(abc)^{\cdot}}{abc} \dot{\phi}
= - \sin{\phi} \left[ m^2_{A }
+\frac{{\cal I} {\cal L}^{\prime}({\cal I})}{\Psi^2_0 \cos{\phi}} \right]\,,
\label{25}
\end{equation}
which shows explicitly that the formula $\phi = \pi n$ with integer $n$ gives  the series of exact solutions to the axion field equation in the case of arbitrary nonlinear term ${\cal L}({\cal I})$ with a finite derivative.

\subsection{Reduced Equations for the Gravity Field}

The equations (\ref{15}) with (\ref{16}) and (\ref{17}) convert into four nonlinear differential equations of the second order
\begin{equation}
\left[\frac{\dot{a}}{a} \frac{\dot{b}}{b} {+} \frac{\dot{a}}{a} \frac{\dot{c}}{c} {+} \frac{\dot{b}}{b} \frac{\dot{c}}{c} \right] = \Lambda  {+} \frac12 \kappa \Psi^2_0
 \left(V{+}{\dot{\phi}}^2 \right)  {+} \kappa {\cal L}(\cal I)\,,
\label{25}
\end{equation}
\begin{equation}
\left[\frac{\ddot{b}}{b} +  \frac{\ddot{c}}{c} + \frac{\dot{b}}{b} \frac{\dot{c}}{c} \right]  = \Lambda  + \frac12 \kappa \Psi^2_0  \left(V-{\dot{\phi}}^2 \right) + \kappa \left[{\cal L}({\cal I}) - 2 {\cal I} {\cal L}^{\prime}({\cal I}) \right] \,,
\label{26}
\end{equation}
\begin{equation}
\left[\frac{\ddot{a}}{a} +  \frac{\ddot{c}}{c} + \frac{\dot{a}}{a} \frac{\dot{c}}{c}\right]  = \Lambda + \frac12 \kappa \Psi^2_0  \left(V-{\dot{\phi}}^2 \right) +\kappa \left[{\cal L}({\cal I}) - 2 {\cal I} {\cal L}^{\prime}({\cal I}) \right]\,,
\label{27}
\end{equation}
\begin{equation}
\left[\frac{\ddot{b}}{b} +  \frac{\ddot{a}}{a} + \frac{\dot{b}}{b} \frac{\dot{a}}{a}\right]  = \Lambda +  \frac12 \kappa \Psi^2_0  \left(V-{\dot{\phi}}^2 \right) {+} \kappa {\cal L}(\cal I)  \,.
\label{28}
\end{equation}
Since only the magnetic and electric fields directed along the axis $0z$ are the sources of the spacetime anisotropy, we can consider the truncated model with the so-called
local isotropy, when $a(t)=b(t)$. Also we introduce the function  $H=\frac{\dot{a}}{a}$, and can rewrite the independent gravity field equations  as the following set:
\begin{equation}
H^2 {+} 2H \frac{\dot{c}}{c}  = \Lambda  {+} \frac12 \kappa \Psi^2_0
 \left(V{+}{\dot{\phi}}^2 \right)  {+} \kappa {\cal L}(\cal I)\,,
\label{0257}
\end{equation}
\begin{equation}
\left[\frac{\ddot{a}}{a} +  \frac{\ddot{c}}{c} + \frac{\dot{a}}{a} \frac{\dot{c}}{c}\right]  = \Lambda + \frac12 \kappa \Psi^2_0  \left(V-{\dot{\phi}}^2 \right) +\kappa \left[{\cal L}({\cal I}) - 2 {\cal I} {\cal L}^{\prime}({\cal I}) \right]\,,
\label{277}
\end{equation}
\begin{equation}
2 \dot{H} + 3 H^2  = \Lambda +  \frac12 \kappa \Psi^2_0  \left(V-{\dot{\phi}}^2 \right) {+} \kappa {\cal L}(\cal I)  \,.
\label{287}
\end{equation}
As usual, we see that the equation (\ref{277}) is the differential consequence of (\ref{0257}) and (\ref{287}).

\subsection{Key Equations of the Model}

The difference of the equations (\ref{0257}) and (\ref{287}) yields
\begin{equation}
 H \frac{\dot{c}}{c} -  \dot{H} -  H^2 = \frac12 \kappa \Psi^2_0 {\dot{\phi}}^2   \,,
\label{257}
\end{equation}
and we obtain immediately the scale factor $c(t)$ in quadratures
\begin{equation}
c(t) = c(t_0) \frac{a(t)H(t)}{a(t_0)H(t_0)} \exp{\left\{\frac12 \kappa \Psi^2_0 \int_{t_0}^t \frac{dt}{H(t)} {\dot{\phi}}^2 \right\}} \,.
\label{2570}
\end{equation}
For the model with the local isotropy it is convenient to use the new variable $x$ introduced as follows:
\begin{equation}
x = \frac{a(t)}{a(t_0)} \,, \quad \frac{d}{dt} = xH \frac{d}{dx} \,.
\label{29}
\end{equation}
In these terms the scale factor (\ref{2570}) takes the form
\begin{equation}
 c(x) = c(1) \ x \frac{H(x)}{H(1)} \exp{\left[\frac12 \kappa \Psi^2_0 \int_1^x zdz {\phi^{\prime}}^2 (z) \right]}  \,,
\label{255}
\end{equation}
and we obtain the pair of the {\it key equations} for two unknown functions $H(x)$ and $\phi(x)$:
\begin{equation}
\frac{1}{x^2} \ \frac{d}{dx}(x^3 H^2)  = \Lambda +  \frac12 \kappa \Psi^2_0  \left[2 m^2_A(1-\cos{\phi})- x^2H^2{\phi^{\prime}}^2 \right]  {+} \kappa {\cal L}(\cal I)  \,.
\label{k1}
\end{equation}
\begin{equation}
\frac{1}{x^2} \ \frac{d}{dx}(x^4 H^2 \phi^{\prime}) + \frac12 \kappa \Psi^2_0 x^3 H^2  {\phi^{\prime}}^3 = - \sin{\phi} \left[ m^2_{A }
+\frac{{\cal I} {\cal L}^{\prime}({\cal I})}{\Psi^2_0 \cos{\phi}} \right]\,.
\label{k2}
\end{equation}
We have to keep in mind now that the argument ${\cal I}$ of the nonlinear Lagrangian can be written as
\begin{equation}
{\cal I} = \frac{\omega^2}{x^4 \cos{\phi}} \,, \quad \omega^2 \equiv \frac{B_0^2}{2a^4(t_0)} \,,
\label{k3}
\end{equation}
and that the scale factor $a(t)$ can be reconstructed using the formula
\begin{equation}
 t-t_0 = \int^{\frac{a(t)}{a(t_0)}}_1 \frac{dx}{x  H(x)} \,.
\label{k4}
\end{equation}

\section{Analysis of the Model: Exact Solutions and Numerical Study}

The plan of investigations of the presented model is the following. First of all, we consider the model with ${\cal L}({\cal I}) = {\cal I}$, as an intermediate case, i.e., the axion field is presented in the nonlinear context, but the electrodynamics is linear. The main question which we have to discuss here, can be formulated as follows: are the solutions for the electric field finite or infinite, or equivalently, is the model stable or instable? The second step is to consider three particular nonlinear models, for which the Hubble functions are found analytically, but the axion field and electric field are obtained using numerical calculations.

\subsection{Quasi-Linear Model}

\subsubsection{ The Test of Instability}

 For the model with ${\cal L}({\cal I}) = {\cal I}$ two key equations (\ref{k1}) and (\ref{k2}) can be rewritten in the form:
\begin{equation}
\frac{1}{x^2} \ \frac{d}{dx}(x^3 Y^2)  = \Lambda^{*} +   \kappa \Psi^2_0  \left[ \mu^2_A(1-\cos{\phi})- \frac12 x^2Y^2{\phi^{\prime}}^2   {+}  \frac{1}{x^4 \cos{\phi}} \right] \,.
\label{link1}
\end{equation}
\begin{equation}
\frac{1}{x^2} \ \frac{d}{dx}(x^4 Y^2 \phi^{\prime}) + \frac12 \kappa \Psi^2_0 x^3 Y^2  {\phi^{\prime}}^3 = - \sin{\phi} \left[ \mu^2_{A }
+\frac{1}{x^4 \cos^2{\phi}} \right]\,.
\label{link2}
\end{equation}
In order to simplify the numerical simulation we introduced here the auxiliary quantities $Y(x)$, $\mu_{A}$, $\Lambda^{*}$ given by
\begin{equation}
H(x) = Y(x) H^{*} \,,  \quad m_{A} = \mu_{A} H^{*} \,, \quad \Lambda = \Lambda^{*} {H^{*}}^2 \,, \quad H^{*} \equiv \frac{|B_0|}{\sqrt2 a^2(t_0)\Psi_0}
\,.
\label{aux1}
\end{equation}
Similarly, we can introduce the dimensionless electric field $e(x)$ as follows:
\begin{equation}
E(x) = e(x) E^{*} \,, \quad E^{*}  \equiv \frac{|B_0|}{a^2(t_0)} \ \ \rightarrow \ \ e(x) = \frac{\tan{\phi(x)}}{x^2}
\,.
\label{aux2}
\end{equation}
When $\phi \to \frac{\pi}{2}$, the electric field (\ref{23}), the invariants of the electromagnetic fields (\ref{42}) and the basic invariant (\ref{43}) tend to infinity. Respectively, the right-hand sides of the key equations (\ref{link1}) and (\ref{link2}) also tend to infinity in this limit. We have to test, whether the axion field $\phi$ can reach the catastrophic value $\phi= \frac{\pi}{2}$.
For this purpose we consider two submodels. In the first one the initial value of the axion field is chosen as $\phi(t_0)=0$; this value corresponds to the first minimum of the axion potential (\ref{14}). In the second submodel we assume that the axion field evolution starts with the value $\phi=\pi$, which relates to the first maximum of the axion potential.

\begin{figure}
\includegraphics[width=10.5 cm]{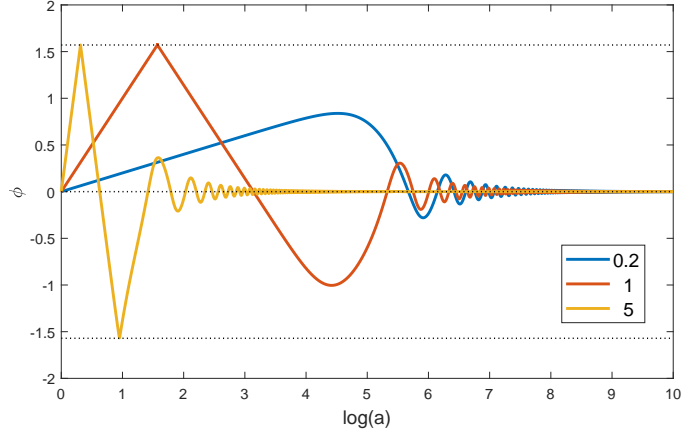}
\caption{This figure illustrates the behavior of the axion field with the initial value $\phi(t_0)=0$ as a function of the initial value of the axion field derivative $\dot{\phi}(t_0)$; three values of this parameter are fixed in the lower right corner of the panel. All the alues of the axion field $\phi$ belong to the interval $-\frac{\pi}{2}< \phi < \frac{\pi}{2}$. The character of evolution of the axion field can be characterized as damping oscillations. \label{fig1}}
\end{figure}

\begin{figure}
\includegraphics[width=10.5 cm]{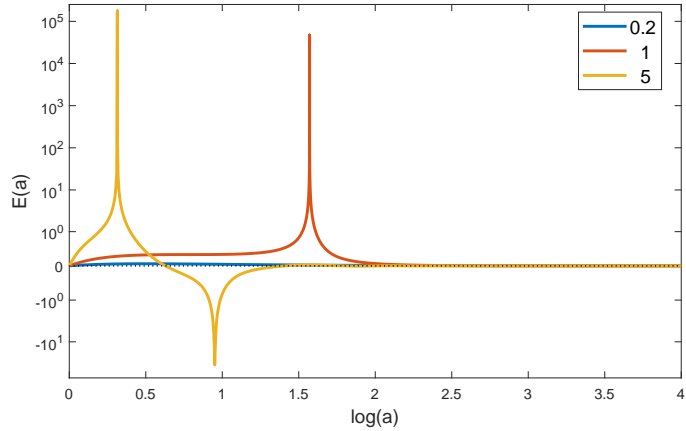}
\caption{This figure presents three graphs of the reduced electric field $e(x)= \frac{E(x)}{E^{*}}$, which relate to the graphs of $\phi$ depicted on Fig.1. ($E^{*}= \frac{|B_0|}{a^2(t_0)}$). The electric field reaches very big values for the time moments, when the values of the axion field are close to $\pm \frac{\pi}{2}$. The behavior of the electric field inherits the oscillatory regime of the axion field, however, this property is hidden on the graph because of the incommensurability of the amplitudes of the first and second maxima of the corresponding graphs. \label{fig2}}
\end{figure}

\begin{figure}
\includegraphics[width=15 cm]{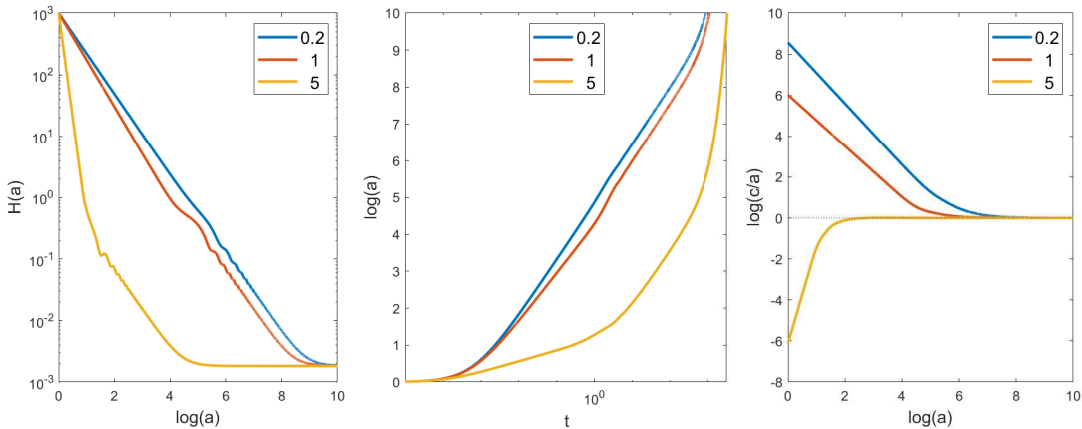}
\caption{Illustrations of the geometric properties of the model. The behavior of the reduced Hubble function $Y(x)= \frac{H(x)}{H^{*}}$ is presented on the left panel; the central panel contains the graph of the scale factor $a(t)$; the right panel presents the evolution of the ratio $\frac{c(t)}{a(t)}$. Oscillations of the axion field declare themselves in the form of ripples on the graph of the Hubble function. Asymptotically, $H(x)$ tends to the de Sitter constant $\sqrt{\frac{\Lambda}{3}}$, and the ratio $\frac{c(t)}{a(t)}$ tends to one thus confirming the fact of the Universe isotropization. \label{fig3}}
\end{figure}

\subsubsection{The First Submodel is stable}

When $\dot{\phi}(t_0)=0$, we see that $\phi(x) \equiv 0$ is the exact solution to the equation (\ref{link2}). As for the equation (\ref{link1}), it can be directly integrated yielding
\begin{equation}
Y^2(x) = \frac{1}{x^3} Y^2(1) + \frac13 \Lambda^{*}\left(1- \frac{1}{x^3} \right) + \kappa \Psi^2_0 \left(\frac{1}{x^3}-\frac{1}{x^4} \right)  \,.
\label{link11}
\end{equation}
Since $E(x)=0$ for this solution, we recover the well known solution attributed to the Bianchi-I model with the pure magnetic field. Asymptotically, this solution describes de Sitter model with $H \to \sqrt{\frac{\Lambda}{3}}$, and is characterized by isotropization $\frac{c}{a} \to const$.

When $\dot{\phi}(t_0) \neq 0$, the numerical analysis of the key equations (\ref{link1}), (\ref{link2}) has shown that all the curves $\phi(x)$ are located below the horizontal line $\phi = \frac{\pi}{2}$. This means that the electric field $E$, Hubble function $H$ and scale factors $a$, $c$ take finite values, i.e., the catastrophic regime can not be realized. We illustrate this fact using Fig.1.-Fig.3

For the model under discussion the oscillatory regime of the axion field evolution is predetermined by the structure of the equation (\ref{link2}). Indeed, in the vicinities of the extrema points, when $\phi^{\prime} \to 0$, the mentioned equation can be rewritten as
\begin{equation}
\phi^{\prime \prime}  = -  \frac{1}{x^{2} Y^{2}} \left( \mu^2_{A} +\frac{1}{x^4 \cos^2{\phi}} \right) \sin{\phi} \,.
\label{link24}
\end{equation}
Clearly, the sign of the second derivative depends on the sign of the function $\sin{\phi}$:  when $\phi>0$, we see that $\phi^{\prime \prime}<0$, and we observe the maxima; when $\phi<0$, we deal with the minima. If $\phi \to \frac{\pi}{2}$, the corresponding second derivative becomes large due to the term proportional to $\frac{1}{\cos^2{\phi}}$, this means that the degree of convexity/concavity of the graph increases, but the typical structure of the extrema is preserved. The damping of these oscillations is provided by the multipliers of the type $\frac{1}{x^4}$ and $\frac{1}{x^2}$, which decrease when the reduced scale factor $x=\frac{a(t)}{a(t_0)}$ grows.

\subsubsection{The Second Submodel is Instable}

When we put $\phi(t_0)=\pi$ and $\dot{\phi}(t_0) =0$, we again obtain the constant exact solution $\phi \equiv \pi$ to the equation (\ref{link2}).
The Hubble function is now described by the formula
\begin{equation}
Y^2(x) = \frac{1}{x^3} Y^2(1) + \frac13 \tilde{\Lambda} \left(1- \frac{1}{x^3} \right) - \kappa \Psi^2_0 \left(\frac{1}{x^3}-\frac{1}{x^4} \right)  \,,
\label{link11}
\end{equation}
where the redefined cosmological constant is $\tilde{\Lambda} = \Lambda^{*} + 2\kappa \Psi^2_0 \mu^2_{A }$. Again, at $x \to \infty$ we obtain the de Sitter type behavior of the model.

When $\dot{\phi}(t_0) \neq 0$, the numerical analysis of the key equations (\ref{link1}), (\ref{link2}) gives principally different results: the horizontal line $\phi {=} \frac{\pi}{2}$ happens to be crossed for arbitrary nonvanishing initial value of the derivative $\dot{\phi}(t_0)$. This submodel is instable. We illustrate this fact by the Fig.4.

\begin{figure}
\includegraphics[width=10.5 cm]{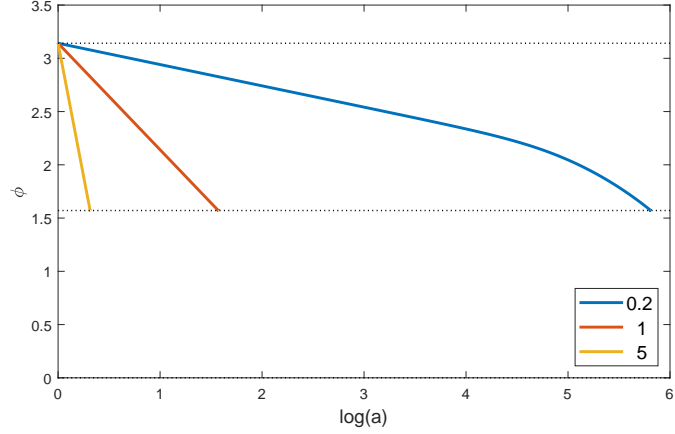}
\caption{Illustrations of the behavior of the axion field with initial value $\phi(t_0){=}\pi$. The graphs are monotonic, and the axion field reaches the value $\phi = \frac{\pi}{2}$ for the finite value of the reduced scale factor $x$. The corresponding values of the electric field are infinite. The geometric characteristics of the Universe reveal singularity. \label{fig4}}
\end{figure}

\subsection{Analysis of the Behavior of Nonlinear Systems in the Vicinity of the First Minimum of the Axion Potential: Three Explicit Examples}

\subsubsection{Basic Equilibrium Solutions and Equation of Perturbation Dynamics}

The first minimum of the axion potential corresponds to the value $\phi=0$; we add to the axion field a small spatially homogeneous perturbation $\psi(t)$. In the leading order approximation the electric field (\ref{23}) vanishes, and the key equation (\ref{k1}) takes the form
\begin{equation}
\frac{1}{x^2} \ \frac{d}{dx}(x^3 H^2)  = \Lambda +  \kappa {\cal L}(\cal I)  \,.
\label{k19}
\end{equation}
Since now the invariant ${\cal I}$ is of the form ${\cal I} \to \frac{\omega^2}{x^4} \equiv z$, we can rewrite the key equation as follows:
\begin{equation}
-4z \frac{d H^2}{dz} + 3H^2 = \Lambda +  \kappa {\cal L}(z) \,.
\label{319}
\end{equation}
The solution for $H(x)$ can be presented in quadratures
\begin{equation}
H^2(x) = H^2(1) x^{-3}+ \frac{\Lambda}{3}(1-x^{-3}) - \frac14 \kappa \omega^{\frac32} x^{-3} \int_{\omega^2}^{\omega^2 x^{-4}} d\xi \ \xi^{-\frac74} {\cal L}(\xi) \,.
\label{32}
\end{equation}
The scale factor $c(x)$ also can be simplified:
\begin{equation}
 c(x) = c(1) \ x \frac{H(x)}{H(1)}  \,.
\label{244}
\end{equation}
In the first order approximation with respect to $\psi(t)$ we obtain the equation
\begin{equation}
\frac{1}{x^2} \ \frac{d}{dx}\left[x^4 H^2(x) \frac{d}{dx} \psi \right] + \psi \left[ m^2_{A }
+\frac{\omega^2{\cal L}^{\prime}(z)}{x^4 \Psi^2_0} \right] =0 \,.
\label{k2q}
\end{equation}
In the linear approximation the electric field is proportional to $\psi$, i.e., $E(x)= \frac{\psi}{x^2} E^{*}$.

\subsubsection{The First Example: Power-Law Lagrange Function}

Integration in (\ref{32}) with the function (\ref{L1}) gives
$$
H^2(x) = H^2(1) x^{-3}+ \frac{\Lambda}{3}(1-x^{-3}) +
$$
\begin{equation}
+ \kappa \left\{\frac{\omega^2}{x^4}(x-1) + \frac{\gamma}{(4\nu -3)} \left(\frac{\omega^2}{x^4} \right)^{\nu} \left(x^{4\nu-3}-1 \right)\right\}\,.
\label{L101}
\end{equation}
Asymptotically, we deal with the de Sitter law $H(x \to \infty) \to \sqrt{\frac{\Lambda}{3}}$.

\begin{figure}
\includegraphics[width=15 cm]{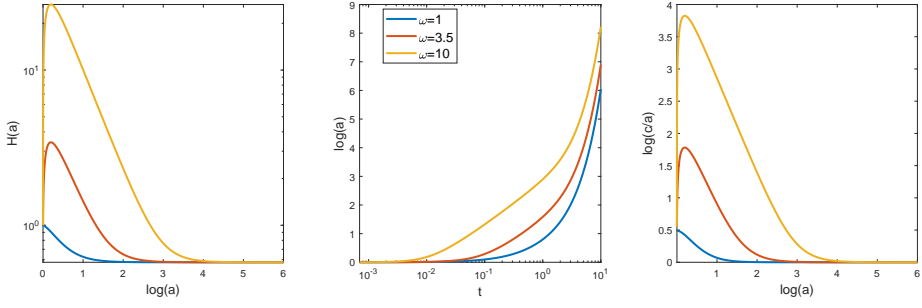}
\caption{Illustration of the behavior of the Hubble function $H(x)$, of the scale factor $a(t)$ and of the ratio $\frac{c(t)}{a(t)}$ for a few values of the parameters $\nu$ and $\omega$ 
attributed to the Power-Law Lagrange function. \label{fig5}}
\end{figure}

\begin{figure}
\includegraphics[width=10.5 cm]{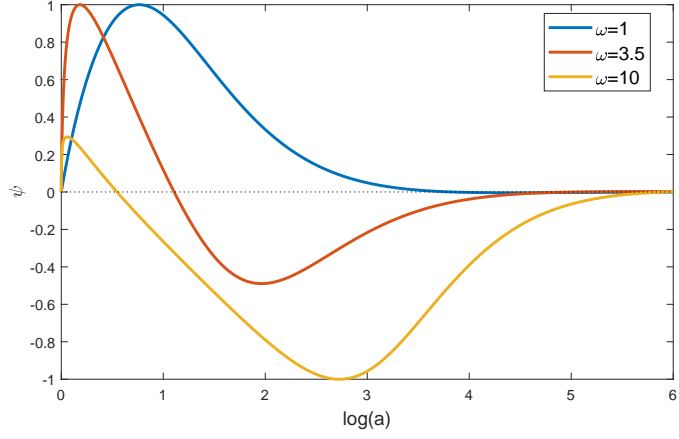}
\caption{Illustration of the behavior of the function $\psi$ for the model with Power-Law Lagrange function. Perturbations remain finite, the model is stable. \label{fig6}}
\end{figure}

\subsubsection{The Second Example: Kohlrausch Type Function with $\gamma = \frac18$}

Now we consider the Lagrange function (\ref{L3}) with $\gamma {=} \frac18$. The function $H(x)$ can be presented in the explicit analytic form
$$
H^2(x) = H^2(1) x^{-3}+ \frac{\Lambda}{3}(1-x^{-3}) +
$$
\begin{equation}
+ 2\kappa \frac{\omega^3}{x^6}\left\{\omega^{\frac14}\left(\frac{1}{\sqrt{x}}{-}1 \right) {+} \exp{\left(\omega^{\frac14}\right)} {-} \exp{\left[\left(\frac{\omega}{x^2}\right)^{\frac14}\right]}   \right\}\,.
\label{L301}
\end{equation}
Again, we deal with the asymptotically de Sitter type behavior, however, in contrast to the results for the quasi-linear model, one can see that $H(x)$ and $\frac{c(t)}{a(t)}$ are non-monotonic.

\begin{figure}
\includegraphics[width=15 cm]{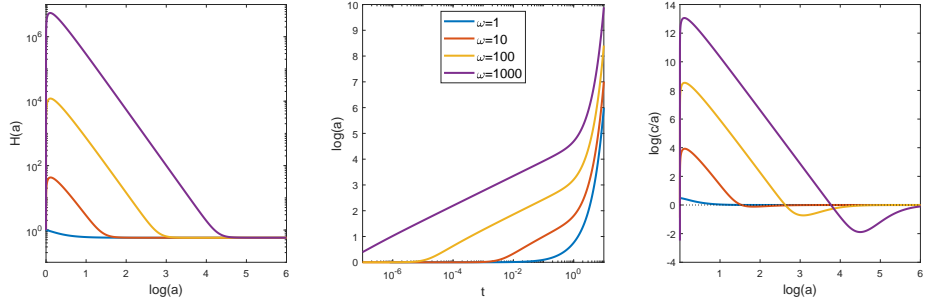}
\caption{Illustration of the behavior of the function $H(x)$, $a(t)$ and of the ratio $\frac{c(t)}{a(t)}$ for the Kohlrausch model with $\gamma = \frac18$.  \label{fig7}}
\end{figure}

\begin{figure}
\includegraphics[width=10.5 cm]{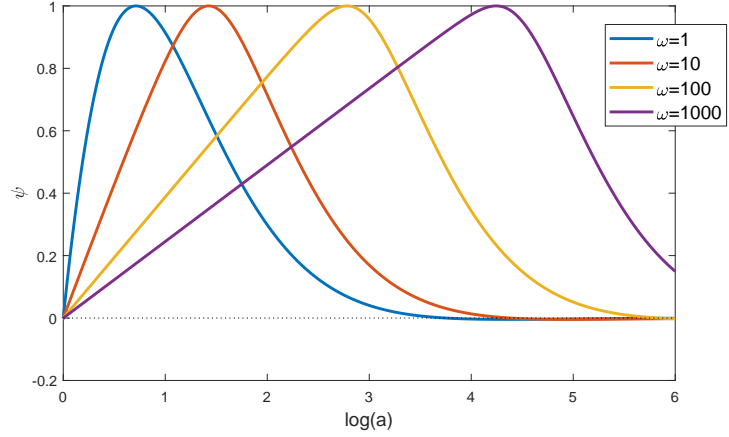}
\caption{Illustration of the behavior of the function $\psi$ for the Kohlrausch model with $\gamma = \frac18$. Perturbations of the axion field remain finite and their moduli do not exceed the value $\frac{\pi}{2}$. \label{fig8}}
\end{figure}

\begin{figure}
\includegraphics[width=15 cm]{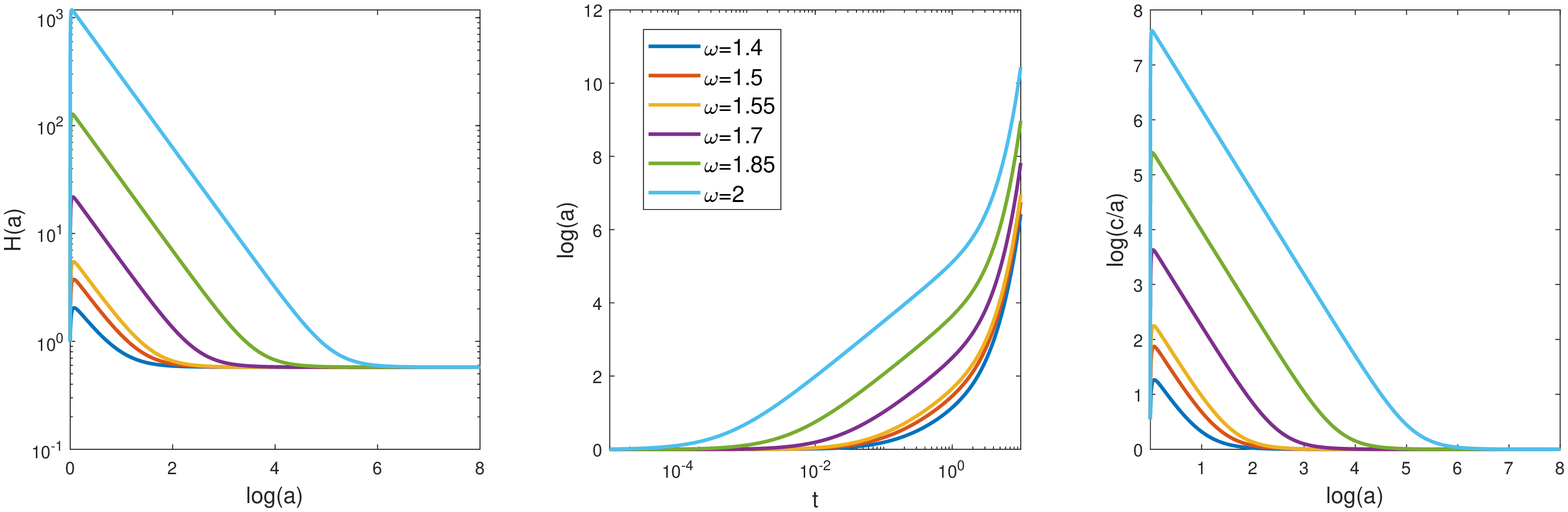}
\caption{Illustration of the behavior of the functions $H(x)$, $a(t)$ and $\frac{c(t)}{a(t)}$ for the Anti-Gaussian Lagrange function (\ref{L302}). \label{fig9}}
\end{figure}

\begin{figure}
\includegraphics[width=10.5 cm]{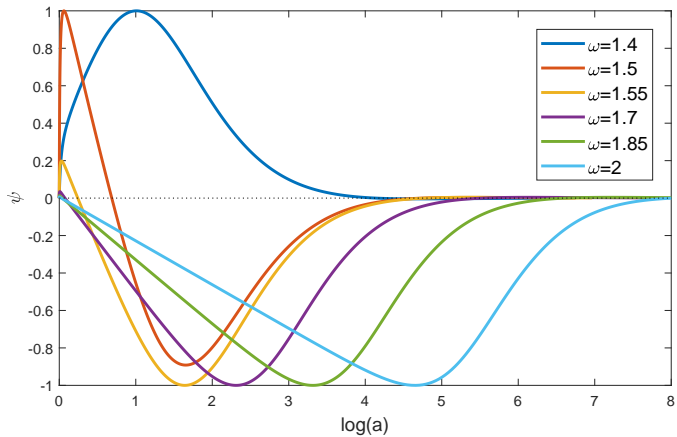}
\caption{Illustration of the behavior of the function $\psi(t)$ for the Anti-Gaussian Lagrange function (\ref{L302}).\label{fig10}}
\end{figure}

\subsubsection{The Third Example: Kohlrausch Type Function with $\gamma {=} 2$ (Anti-Gaussian Function)}

When we consider the model Lagrangian of the form
\begin{equation}
{\cal L}({\cal I}) = {\cal I} + \beta {\cal I}^{\frac{11}{4}} \left[e^{{\cal I}^2} -1 \right] \,,
\label{L302}
\end{equation}
we obtain explicit analytic formula for the Hubble function; its square is
$$
H^2(x) = H^2(1) x^{-3}+ \frac{\Lambda}{3}(1-x^{-3}) +
$$
\begin{equation}
  +\kappa \left\{\frac{\omega^2}{x^4}(x{-}1) {+} \frac{\beta}{8} \left[\left(\frac{\omega^2}{x^4} \right)^{\frac{11}{4}}(1{-}x^8) {+} \left(\frac{\omega^2}{x^4} \right)^{\frac{3}{4}}\left(e^{\omega^4}{-}e^{\frac{\omega^4}{x^8}} \right) \right]\right\} \,.
\label{32v}
\end{equation}
Fig.9 illustrates the behavior of $H(x)$, of the scale factor $a(t)$ and of the ratio $\frac{c(t)}{a(t)}$ for this model.
Fig.10. presents the corresponding graphs of the function $\psi(t)$.

\section{Discussion and Conclusions}

\subsection{Axionically Induced Electric Field: Anomalies and Electric Flares}

The mechanism of generation of the axionically induced electric field in the presence of the magnetic field is well known and well documented. This mechanism is based on the phenomenon of axion-photon coupling, and it has a lot of interesting applications (see, e.g., \cite{4,5,M7}). The new trend in the axion electrodynamics is connected with the elaboration of the  Nonlinear Axion Electrodynamics, which could have astrophysical and cosmological applications. In this sense, the early Universe with strong magnetic field, filled with the axionic dark matter \cite{DM1,DM2,DM3}, seems to be one of the  most interesting objects of research. Indeed, it is well-known that the dark matter axions interacting with the cosmic magnetic field produce the electric field, however, as the linear axion electrodynamics predicts, the contribution of such electric field into the energetic balance of the Universe seems to be negligible, since the coupling constant of the axion-photon interactions is rather small, $\frac{1}{\Psi_0}=g_{A \gamma \gamma} < 1.47 \cdot 10^{-10} {\rm GeV}^{-1}$. The situation changes principally, when we work with Nonlinear Axion Electrodynamics. As we have shown above, the electric field with zero initial value can grow proportionally to the function $\tan{\phi}$. This means that if the dimensionless function $\phi$, describing the evolution of the axion field, tends to the value $\phi {=} \frac{\pi}{2}$, the axionically induced electric field grows abnormally. We have studied the special question: is the catastrophic value $\phi {=} \frac{\pi}{2}$  reachable? The answer is negative, however, the maximum value of the electric field can be very large and even huge, depending on the initial value of the derivative $\dot{\phi}(t_0)$.
As it was shown above, the anomalous behavior of the electric field can be indicated as flare, since this electric field rapidly decreases in amplitude; this phenomenon can occur only once in the early Universe.
If the initial value of the axion field is equal to $\phi(t_0)=\pi$, i.e., if it corresponds to the first maximum of the axion field potential, the catastrophic value $\phi {=} \frac{\pi}{2}$ will be achieved inevitably; such model happens to be singular.

What are the physical consequences of such electric flares? From our point of view one can expect, first, the creation of the electron-positron pairs, second, the acceleration of charged particles, third, the emission of the electromagnetic waves by these accelerated particles.

\subsection{On the Electron-Positron Pair Creation in the Axionically Induced Electric Field}

According to the Schwinger's formula, the probability of the electron-positron pair creation in vacuum under the influence of a strong electric field $E$ is estimated to be the following:
\begin{equation}
{\cal W}=\frac{e^{2} E^{2}}{4 \pi^{3} \hbar^{2}} \exp{\left(-\frac{E_{(crit)}}{E}\right)} \,, \quad  E_{(crit)}=\frac{\pi m_{e}^{2}}{\hbar e} \,, \quad (c=1)\,.
\label{Schw1}
\end{equation}
Clearly, the Nonlinear Axion Electrodynamics of the early magnetized Universe does not prohibit the anomalous electric field energy to overcome the threshold $2m_{e}c^2$, thus providing the creation of the electron-positron pairs in vacuum.

\subsection{Particle Acceleration}

If the initial three-momenta of the electron and positron, born in the strong electric field, are equal to zero, these particles begin to move along the direction $0z$ indicated by the parallel electric and magnetic fields. For this field configuration the particles will have only two components of the velocity four-vector: $U_z$ and $U_0=\sqrt{1+c^{-2}(t)U^2_z}$, and the longitudinal component of the acceleration four-vector. From the equation of the particle dynamics
\begin{equation}
\frac{d P_j}{ds} = \Gamma^k_{jl}P_k \frac{P^l}{m_e} + \frac{e}{m_e} F_{jl}P^l \,,
\label{Schw2}
\end{equation}
where $P_i = m_e U_i$ is the particle four-momentum, we obtain for the longitudinal acceleration and velocity, respectively, the following relationships:
\begin{equation}
\frac{d U_z}{dt} = - \frac{e c(t)B_0}{m_e a^2(t)} \ \tan{\phi(t)} \,, \quad U_z(t) = - \frac{e B_0}{m_e} \int^t_{t_0} dt' \frac{c(t')}{a^2(t')} \tan{\phi(t')} \,.
 \label{Schw2}
\end{equation}
These kinematic characteristics depend on time and thus we deal with the non-uniform accelerated motion. Clearly, electrons and positrons move in opposite directions and can take part in the process of anomalous acceleration, if $\phi(t) \to \frac{\pi}{2}$.

\subsection{On the Radiation of Electromagnetic Waves by the Charged Particles Accelerated by the Axionically Induced Electric Field}

The non-uniformly moving particles emit the electromagnetic wave. We consider the radiation of particles, which have the parallel three-velocity $\vec{v}$ and three-acceleration $\vec{w}$. In this case the standard formulas for the radiation intensity per single solid angle has the form
\begin{equation}
\frac{d I}{d\Omega} =  \frac{e^2}{4\pi} \frac{w^2 \sin^2{\theta}}{\left(1- v \cos{\theta}\right)^6} \,, \quad v=|\vec{v}| \,, \quad w = |\vec{w}| \,, \quad (c=1) \,,
 \label{Schw3}
\end{equation}
were $\theta$ is the angle between the axis $0z$ and the direction to the observation point (see, e.g., \cite{LL}). In our case for the very sharp peak of the electric flare we can use the formulas $v \to \left|\frac{U_z}{U_0}\right|$ and $w \to \frac{d U_z}{dt}$ (see (\ref{Schw2})).  In the case of the anomalous axionically induced electric field the charged particle born in this field can reach the ultrarelativistic state of motion, and the radiation will be focused in the small angle $\theta$. The intensity of radiation is proportional to $B^2_0 \tan^2{\phi}$. In other words, one can expect that the directions $\theta=0$ and $\theta =\pi$ will be marked by the extra portions of light.

\subsection{Conclusions}

We have presented the new version of the nonlinear extension of the Einstein-Maxwell-axion theory, and based on this theory we can propose the following scenario of the evolution of the spatially homogeneous early magnetized Universe.

\noindent
1. Let the axion field evolution start with zero value $\phi(t_0)=0$ and with non-vanishing derivative $\dot{\phi}(t_0)\neq 0$; in the presence of the cosmological magnetic field the axion-photon coupling provides the creation of the axionically induced electric field parallel to the magnetic field; in its turn, such electromagnetic field configuration produces a new portion of axions, thus unwinding the spiral of the nonlinear axion-photon interactions.

\noindent
2. In the process of evolution the axion field and its derivative remain finite; the axion field itself happens to be trapped in the zone $-\frac{\pi}{2} < \phi < \frac{\pi}{2}$, and thus does not play essential role in the Universe expansion.

\noindent
3. The axionically induced electric field, in contrast to the axion field, can reach anomalously large values, when $\phi \to \frac{\pi}{2}$; the so-called electric flares can appear, and the electromagnetic contribution to the sources of the gravity field can become essential.

\noindent
4. The geometric characteristics of the Universe, such as scale factors and their derivatives, remain nonsingular at $t>t_0$ and inherit the oscillatory behavior of the electromagnetic sources; when the cosmological constant is non-vanishing, the final stage of the Universe evolution is of the de Sitter type, and thus corresponds to the instruction about the late-time accelerated expansion.

\noindent
5. The axionically induced anomalous electric field produces the electron-positron pairs and accelerates the born charged particles; the accelerated charged particles emit the electromagnetic waves, which are focused in the direction pointed by the parallel magnetic and electric fields; these flares can take part in the formation of fluctuations of the cosmic microwave background.

In the next work we hope to revive this scenario by adding estimations and other necessary details.

\acknowledgments{The work was supported by Russian Foundation for Basic Research (Grant N 20-52-05009)


\begin{thebibliography}{999}

\bibitem{PQ} Peccei, R.D.; Quinn, H.R. CP conservation in the presence of instantons. {\em Phys. Rev. Lett.} {\bf 1977}, {\em 38}, 1440--1443.

\bibitem{1} Weinberg, S. A new light boson? {\em Phys. Rev. Lett.} {\bf 1978}, {\em 40}, 223--226.

\bibitem{2} Wilczek, F. Problem of strong P and T invariance in the presence of instantons. {\em Phys. Rev. Lett.} {\bf 1978}, {\em 40}, 279--282.

\bibitem{3} Ni, W.-T. Equivalence principles and electromagnetism. {\em Phys. Rev. Lett.} {\bf 1977}, {\em 38},  301--304.

\bibitem{4} Sikivie, P. Experimental tests of the ``invisible'' axion. {\em Phys. Rev. Lett.} {\bf 1983}, {\em  51}, 1415--1417.

\bibitem{5} Wilczek, F.  Two applications of axion electrodynamics. {\em Phys. Rev. Lett.} {\bf 1987}, {\em 58},  1799--1802.

\bibitem{Jackson}  Jackson, J.D. {\em Classical Electrodynamics}, John Wiley and Sons, USA, 1999.

\bibitem{BaGa} Balakin, A.B.; Galimova, A.A. Towards nonlinear axion-dilaton electrodynamics: How can axionic dark matter mimic dilaton-photon interactions? {\em Phys. Rev. D} {\bf 2021}, {\em 104},
044059.

\bibitem{Kohl} Kohlrausch, R. Theorie des elektrischen Ruckstandes in der Leidner Flasche. {\em Ann. Phys. Chem.} {\bf 1854}, {\em 91}, 56--82.

\bibitem{M1} Turner, M.S.;  Widrow, L.M. Inflation-produced, large-scale magnetic fields. {\em Phys. Rev. D} {\bf 1988}, {\em 37}, {2743--2769}.

\bibitem{M2} Grasso, D.;  Rubinstein, H.R. Magnetic fields in the early Universe. {\em Phys. Rept.} {\bf 2001} {\em 348}, 163-266.

\bibitem{M3} Lee, D.-S.; Lee, W.; Ng, K.-W.   Primordial magnetic fields from dark energy. {\em Phys. Lett. B} {\bf 2002}, {\em 542}, 1--7.

\bibitem{M4} Bronnikov, K.A.; Chudaeva, E.N.; Shikin, G.N.  Magneto-dilatonic Bianchi-I cosmology: Isotropization and singularity problems.
{\em Class. Quantum Gravity} {\bf{2004}}, {\em 21}, 3389--3403.

\bibitem{M5} Bamba, K.;  Odintsov, S.D.   Inflation and late-time cosmic acceleration in non-minimal Maxwell-$F(R)$
gravity and the generation of large-scale magnetic fields. {\em J. Cosmol. 	Astropart. Phys.} {\bf 2008}, {\em 04}, 024.

\bibitem{M6} Balakin, A.B. Magnetic relaxation in the Bianchi-I universe. {\em Class. Quantum Gravity} {\bf 2007}, {\em 24}, 5221--5245.

\bibitem{M7}  Balakin, A.B.; Muharlyamov, R.K.; Zayats, A.E. Axion-induced oscillations of cooperative electric field in a~cosmic magneto-active plasma. {\em Eur. Phys. J. D} {\bf 2014}, {\em 68}, 159.

\bibitem{DM1}  Duffy, L.D.;  van Bibber, K. Axions as dark matter particles. {\em New J. Phys.} {\bf 2009}, {\em 11}, 1050088.

\bibitem{DM2} Steffen, F.D. Dark Matter candidates---Axions, neutralinos, gravitinos, and axinos.
{\em Eur. Phys. J. C} {\bf 2009}, {\em 59}, 557--588.

\bibitem{DM3} Marsh, D.J.E. Axion Cosmology. {\em Phys. Rept.} {\bf 2016}, {\em 643}, 1-79.

\bibitem{LL} Landau, L.D.; Lifshitz, E.M. {\em The Classical Theory of Field}, Pergamon Press, Oxford, 1971.



\end{thebibliography}
\end{document}